**Energy Repartition and Entropy Generation across the Earth's Bow Shock: MMS observations**

O.V. Agapitov[1,2], V. Krasnoselskikh[2,3], M. Balikhin[4], J. W. Bonnell[2], F.S. Mozer[2], L. Avanov[5]

ABSTRACT

The evolution of plasma entropy and the process of plasma energy re-distribution at the collisionless plasma shock front are evaluated based on the high temporal resolution data from the four MMS spacecraft during the crossing of the terrestrial bow shock. The ion distribution function has been separated into the populations with different characteristic behavior in the vicinity of the shock: the upstream core population, the reflected ions; the gyrating ions; the ions trapped in the vicinity of the shock, and the downstream core population. The values of ion and electron moments (density, bulk velocity, and temperature) have been determined separately for these populations. It was shown that the solar wind core population bulk velocity slows down mainly in the ramp with the electrostatic potential increase but not in the foot region as it was supposed. The reflected ion population determine the foot region properties, so that the proton temperature peak in the foot region is the effect of the relative motion of the different ion populations, rather than an actual increase in thermal speed of any of the ion population. The ion entropy evaluated showed a significant increase across the shock: the enhancement of the ion entropy occurs in the foot of the shock front and at the ramp, where the reflected ions are emerging in addition to the upstream solar wind ions, the anisotropy growing to generate the bursts of ion-scales electrostatic waves. The entropy of electrons across the shock doesn't show significant change: electron heating goes almost adiabatically.

**Introduction**

Collisionless shocks (CS) are ubiquitous in space plasma and astrophysical systems. They play an essential role in the interaction of the solar wind with the planets, and they are supposed to play a crucial role in fundamental problems of astrophysics in the acceleration of cosmic rays (Krymskii, 1977; Bell, 1978; Axford, 1977). CS arise in astrophysical systems where a supersonic flow interacts with obstacle or two high-speed flows of different origins penetrate one another. CS form near many astrophysical objects such as supernova remnants, plasma jets, binary objects, and ordinary stars. Despite a great variety of CS in the Universe currently, only CS in the solar system can be probed using *in-situ* observations. Kennel et al. (1985) noticed that plasma density, temperature, and magnetic field in the hot interstellar medium are similar to those in the solar wind, and Mach numbers of supernova shocks at a phase when they accelerate the most cosmic rays are similar to those of solar wind shocks. Thus, the study of terrestrial and heliospheric CS allow one to quantify and constrain the physics in both these local CS, as well as those found in important astrophysical systems.

[1] Corresponding author agapitov@ssl.berkeley.edu
[2] Space Sciences Laboratory, University of California, Berkeley, CA 94720
[3] LPC2E-University of Orleans, Orleans, France
[4] University of Sheffield, Sheffield, UK
[5] NASA GSFC

CS decelerate the flow from supersonic to subsonic and transform part of the kinetic energy of directed flow motion into thermal energy - thermalization. In gas dynamics, this process inevitably involves particle collisions and the corresponding thermalization produces entropy. The scale of the transition region in such a process should be larger than the mean free path of particles. In collisionless plasmas, this would mean unrealistic spatial scales that in the case of the terrestrial bow shock are comparable with the distance from the Sun to Jupiter. The Earth bow shock and other planetary and interplanetary CS are much thinner than this, and so the physics that determines their thickness is quite different than that of collisional shocks. The actual structure of a quasi-perpendicular CS is more complicated than an abrupt transition from upstream to downstream conditions (Krasnoselskikh et al. 2016). The conventional notion of the quasi-perpendicular Earth's bow shock (the CS standing in the magnetosphere frame) includes several elements: the foreshock region, the foot, the ramp region, the overshoot and undershoot regions, and the magnetosheath region separating the magnetopause from the shock. For supercritical CS (with the Mach number typically greater than 3), the ion reflection presumably determines the main energy conversion (Schwartz et al., 2021, 2022), so the shock rump and foot with the major change of the parameters is of the special importance and it is widely considered as the shock front (Hanson et al. 2020).

The simplified MHD description of CS determines the main asymptotic characteristics (density, velocity, pressure, temperature, etc.) of the downstream flow from the known parameters of the upstream flow through conservation of mass, momentum, energy flux, and equation of state across a "black box" transition region (see the following for details: Kennel; Edmontson; Hada). This model assumes that all the relaxation processes occur inside this transition, and the downstream flow is in thermal equilibrium defined by Rankine-Hugoniot relations representing conservation laws for stationary flows.  This model assumes that the modification of the flow after crossing the shock front is supposed to be irreversible, and thus that the shock naturally produces an increase in entropy (Birn et al. 2004; Liu et al 2007). However, even in the magnetosheath region, the asymptotic conditions determined by Rankine-Hugoniot conditions are not fully satisfied – in other words, the state of the plasma may be stable but not obligatorily corresponding to the final downstream thermal equilibrium. Comparing the plasma parameters in the downstream regions of collisional and colisionless shocks, one may conclude that CS cannot provide full thermalization, i.e., the downstream flow cannot reach new thermal equilibrium corresponding to the state with increased entropy**.** According to the Boltzmann equation for collisional plasma the full time derivative of the distribution function $df/dt$ is equal to the collisional term (Stosszahlansatz). This term describes the effects of collisions and leads to the irreversible evolution towards thermodynamic equilibrium. In case of collisionless plasmas, and if we follow the particular region in the phase space in accordance with Lowville's theorem the phase space density remains constant along that region's trajectory through phase space. Therefore if we follow the same population of particles the corresponding entropy should remain constant. Observations from heliospheric low-Mach number shocks have shown that electrons often are not heated much beyond the adiabatic compression and heating associated with the macroscopic structure (change in $|B|$, cross-shock $E$ field, etc.) of the shock (Bame et al. 1979). However, there have been instances where a given CS produces an entropy increase, manifested as super-adiabatic heating of ion and/or electron populations (Scopke et al. 1990; Scopke 1995; Parks et al. 2012; Lindberg et al. 2022). In a MHD description of the plasma, it is supposed that the different populations of particles are in thermal equilibrium. In that case, entropy is a function of the total density and pressure of the plasma (the fluid approximation). In a more general case, the kinetic Boltzmann definition of entropy as defined above can be estimated from plasma distribution functions. For the terrestrial bow shock and other CS, computing the kinetic Boltzmann entropy requires rapid measurements of the detailed distribution functions of particles (Parks et al. 2012; Lindberg et al. 2022). Parks et al. (2012) first reported the dynamics of entropy per particle (density of entropy) and showed that it does increase across the Earth's bow shock based on the data from Cluster. With one-point measurements Parks et al. (2012) proposed the concept of entropy density: $\sum p_i \ln p_i$ estimated at a point in space, where $p_i = f_i \Delta^3 v_i / N$ (i goes for each of velocity bins in the three-dimensional velocity space, and N is the particle number in the space volume) normalized so that $\sum p_i = 1$ (Montgomery & Tidman 1964). This approach allows to estimate the plasma entropy in the velocity space from the particle detectors measurements of the flux in the set of energy channels and the full solid angle. Observations of electron heating at the Earth bow shock show that electrons are in most cases heated to lower temperatures than the ions (Schwartz et al. 1988), with a Mach number dependence that suggests that there is an inverse correlation between Mach number and electron-ion temperature ratio suggesting almost adiabatic heating of electrons (Vink et al. 2014). To explore the source of this correlation, Park et al. (2012) studied the particle energization at a CS in a particle-in-cell (PIC) numerical system and found that the fluctuating electromagnetic fields are necessary for the entropy density creation throughout the downstream region. The detailed processes leading to entropy changes across the shock were then investigated in PIC plasma models of plasma shocks (Yang et al. 2014; Guo et al. 2017, 2018) and reconnection systems (Liang et al. 2019). Guo et al.(2017, 2018) showed that most of the electron heating in low Mach number shocks is adiabatic the irreversible electron heating indicated by entropy enhancement is also can be observed. The two basic ingredients were identified to be necessary for entropy production: (1) a temperature anisotropy, induced by field amplification coupled to adiabatic invariance; and (2) a mechanism to break the particles adiabatic invariance itself (Guo et al. 2017, 2018). To account for this entropy production, early models of CS included the concept of "anomalous collisions", as first formulated by Sagdeev (1966, 1960) and Galeev (1976). According to this concept, the interactions of upstream plasma particles with waves generated by various plasma instabilities within the shock front play the role of collisions in ordinary collisional gas dynamic shocks. The question whether anomalous collisions play a key role at a terrestrial bow and lead to the irreversibility across the shock front remains open. It is worth noting that entropy is presumably the most precise measure in the characterization of the process of transformation of directed energy into thermal, but it is not widely used in practice, and the direct estimation of the entropy change using the unprecedentedly high-time resolution MMS plasma measurements allows to clarify this problem for the terrestrial bow shock.

In this paper we demonstrate the valuable insights gained by applying the concept of entropy density and its changes across the terrestrial bow shock to determine what processes contribute significantly to irreversible changes in the plasma distributions, and where those processes occur within the shock structure. To do that, we first use the high-cadence, high-resolution MMS particle data to distinguish the proton populations with different characteristic dynamics across the Earth's bow shock (core upstream, reflected, core downstream). We then evaluate the partial moments (density, velocity, temperature, entropy density) of these populations to determine which vary and in what ways. We then use those observations, along with the detailed electromagnetic field measurements on MMS to determine the factors responsible for producing entropy in the CS.

**Dynamics of the particles distribution functions during a bow shock crossing by Magnetospheric Multiscale (MMS) on November 2, 2017**

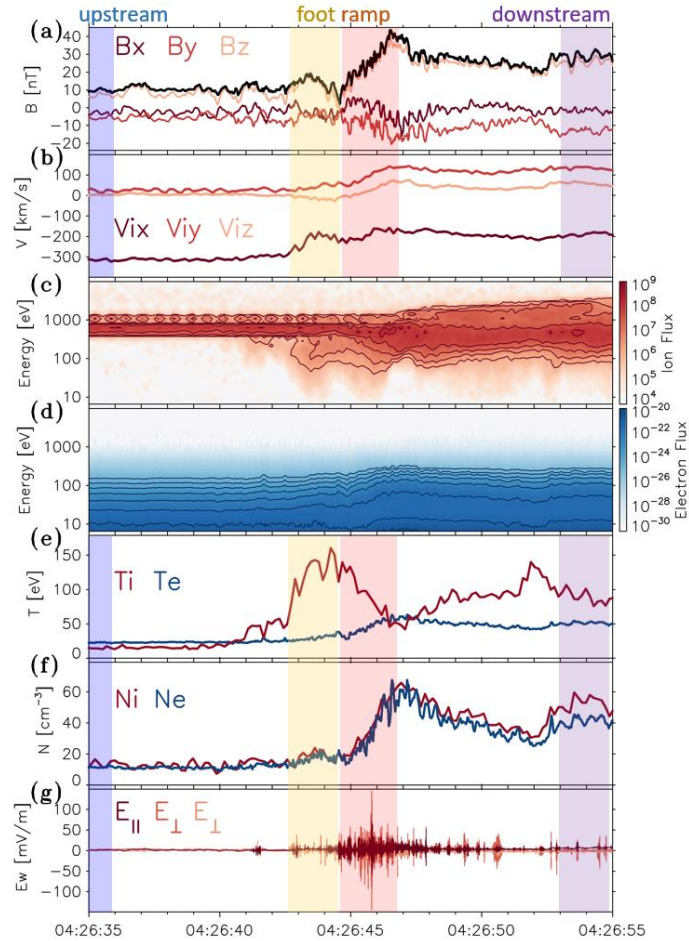

Figure 1. The MMS3 crossing of the bow shock on 2015-11-02: (a) - magnetic field dynamics (the components are presented in the GSE coordinate system; the black curve presents the magnetic field magnitude); (b) – the ion bulk velocity in the GSE coordinate system; (c) the ion omni-directional distribution flux (estimated as the second moment of the entire distribution); (d) – the electron omni-directional flux; (e) - the electron (blue) and ion (red) temperature; (f) - the electron (blue) and ion (red) density; and (g) – the electric perturbations across the shock.

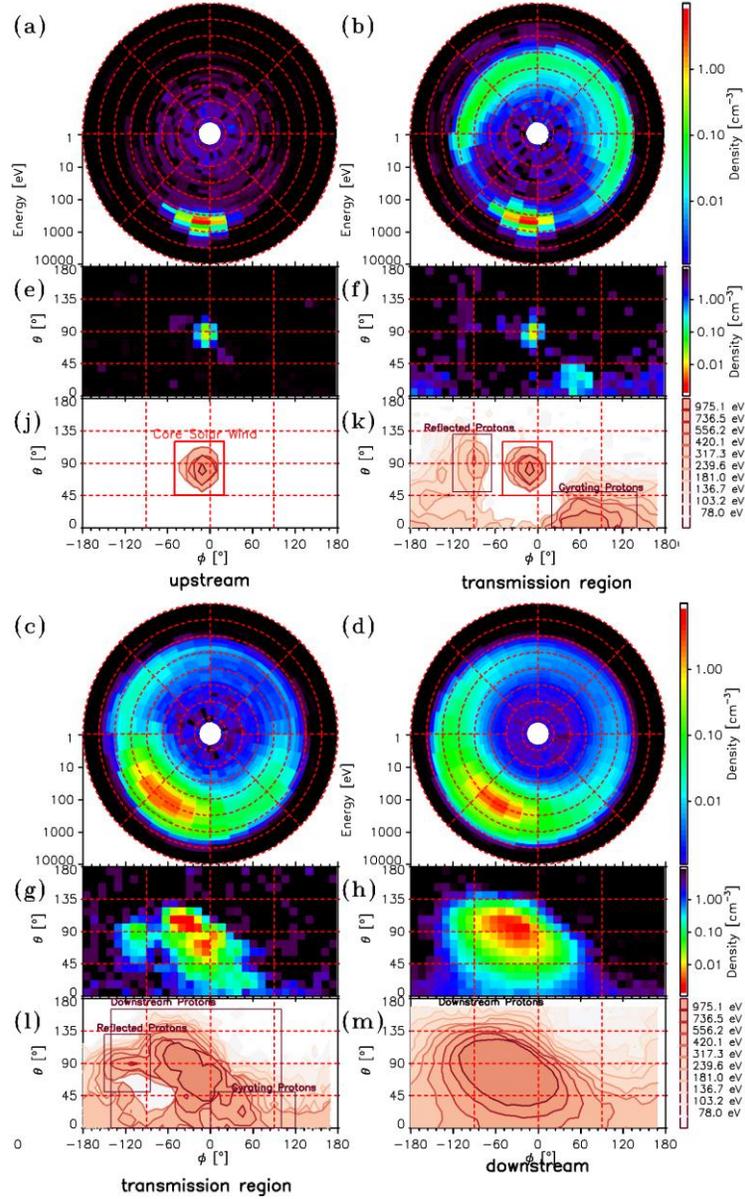

Figure 2. Ion density distributions in the azimuthal angle $\varphi$ ($\varphi = 0$ corresponds to the anti-sunward radial direction) and the ion energy (the radial axis) domain in the far upstream (a), foot (b), ramp (c), and downstream (d) regions. The polar diagrams of the ion phase space density in the $\varphi - \theta$ domain are presented in panels (e, f, g, h) for the same shock regions. The sky diagrams in panels (j,k,l,m) show the directional distribution of the same flux value for the different energies in the $\varphi - \theta$ domain.

Hereafter we discuss a high-beta, quasi-perpendicular CS making use of multipoint observations from the Magnetospheric Multiscale (MMS) constellation (Burch et al. 2013). The MMS mission makes plasma measurements with a high sampling rate and energy/angle resolution that can resolve the dynamics of the particle distribution across the shock front contributed by the different particle populations. MMS is a four-spacecraft constellation in a tetrahedral formation with typical separation distances of tens of km (Fuselier et al. 2016). The Fast Plasma Investigation (FPI) suite aboard MMS spacecraft consists of four dual electrons (DES) and four dual ion (DIS) electrostatic spectrometers (Pollock et al. 2016). The FPI has a high rate of electron and proton distribution function measurements producing the full solid angle flux distribution with 30 S/s for electrons and 3 S/s for protons (Pollock et al. 2016). It provides measurements of the three-dimensional distribution functions in the energy range from 10 eV/q to 30 keV/q on each spacecraft. Using electrostatic field-of-view deflection, the eight spectrometers together provide 4pi-sr field-of-view with 11.25-degree sample spacing. The counts corresponding to each bin in the phase space are combined into a sky map, which is a three-dimensional array with 32 energy, 32 azimuth angles, and 16 polar angles. The FPI continuously operates in burst mode throughout the bow shock crossings. The wave instruments on MMS provide rapid measurements of three components of the electric field (Lindquist et al. 2016; Ergun et al. 2016), three components of the magnetic field from SCM (LeContel et al. 2016), and three components of the background magnetic field from FGM (Russell et al. 2016).

We consider here the bow shock crossing by the MMS spacecraft on November 2, 2015. Figure 1 shows three components of the magnetic field, ion bulk velocity, ion and electron density, and particle fluxes as measured by the MMS3 spacecraft during this shock front crossing. We notice here the effects from helium component, which lead to increase of the estimated proton temperature because of absence of mass resolution in FPI ion measurements (Halekas et al. 2014). The helium component is seen in ion flux (Figure 1c). However, the dynamics of these components is similar across the shock, so we process them together in the following. This bow shock crossing was reported previously by Hanson et al. (2019) and Lindburg et al. (2022). The magnetic field, plasma bulk velocity, and plasma density demonstrate the structure typical for a crossing of a high-Mach number quasi-perpendicular CS. The lower energy ions produce a significant foot region with increased density and magnetic field perturbations as seen in Figure 1.

The ion density distribution across the shock front is shown in Figure 2 for the far upstream (a), foot (b), ramp (c), and downstream (d) plasma regions of the shock. The solar wind flow is highly localized (beamed) in direction and the energy range in the upstream region (Figure 1a,e,j). This focused beamed population remains quite similar in the foot region (highlighted with red in Figure 1a). However, there emerge three additional ion populations, namely, the reflected ions (ions reflected from the shock potential similar to the core solar wind population energy), the gyrating ions accelerated in the ramp region of the shock through drift acceleration (see (Hanson et al. 2020b)), and the downstream ions penetrating to the upstream to the distance about the ion gyroradius (being very similar to the Sun-directed part of the downstream distribution in Figure 2g and Figure 2h). The reflected and downstream populations with different bulk flow velocities significantly complicate the ion plasma distributions through the foot and ramp leading to a biased estimation of plasma bulk velocity and temperature (pressure) (Schwartz et al. 2021, 2022), which makes ion proxies for the evaluation of the shock potential inapplicable (Hanson et al. 2019, 2020a). In the ramp, the solar wind ion core population begins to be heated. One can see this as a spreading of the solar wind beam in Figure 2c,g and an increase of high energy flux in Figure 2l. The downstream distribution shows the effects of the redistribution of solar wind bulk flow kinetic energy to the thermal energy of more dense plasma with the lower bulk speed. As it was described above, because the observed ion populations can be distinguished in phase it is possible to process the parameters of the different populations separately, and more clearly quantify what changes to drift, temperature, and density in different parts of phase space.

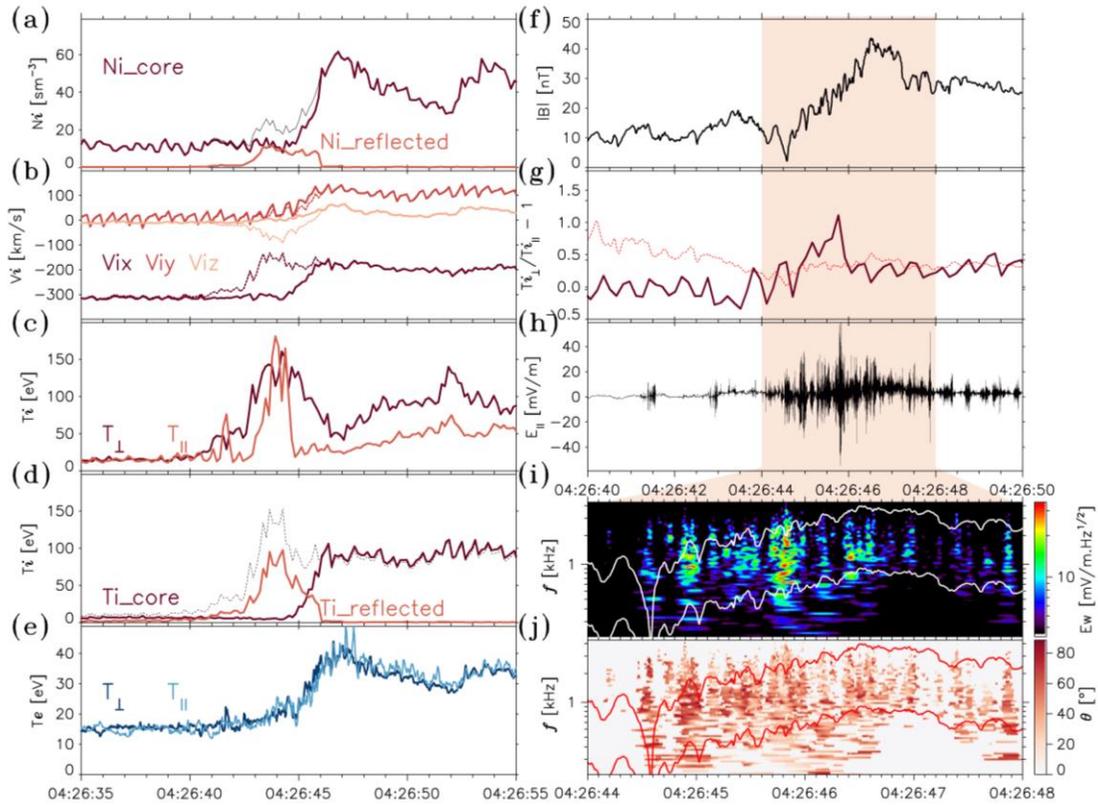

Figure 3. (a) - density of the core ion population and the reflected ion population. (b) - the bulk velocity of core population ions (the solid curves); here and in panels (b) and (d) the estimation made from the entire distribution (Figure 1) is shown with the dashed curves. (c) – ion temperatures estimated from the entire distribution. (d) - temperatures estimated for the core and ion reflected populations. (e) - the electron temperatures: parallel and transverse to the background magnetic field. (f) – the background magnetic field magnitude. (g) – the ion anisotropy $A = T_\perp / T_{||} - 1$ estimated for the core ion population. The red dashed curve shows the anisotropy threshold for the ion cyclotron wave generation. (h) – the parallel component of electric field perturbation (Figure 1). (i) – the dynamic spectrum of the electric field perturbation during the interval highlighted in panels (f-h). (j) – the wave normal angle of the electrostatic perturbations.

Calculation of the partial plasma moments (density, bulk speed, and temperature) of the core upstream/downstream and reflected ion populations shows that the reflected population has density comparable with the core solar wind population in the foot region (Figure 3a). This affects the direction of the bulk velocity in the foot region: Figure 3b presents the revised three components of core ion bulk velocity, and shows that significant changes of the distribution function and slow down of the core population bulk velocity happens only on the ramp of the shock. The parallel and transverse to the background magnetic field temperatures of ions, $Ti_{||}$ and $Ti_\perp$, estimated from the full distribution function are presented in Figure 3c. The core population is heated across the shock ramp and gains anisotropy of about $Ti_\perp / Ti_{||} \sim 1.5$-$1.8$. The electron population is also heated on the shock ramp but much less, from ~$15 \pm 1$ eV to $32 \pm 2$ eV (Figure 3e). Heating of electrons is going isotopically – the parallel and transverse temperatures increase together to the similar values in an agreement with (Schwartz et al. 2011), which reported that the major electron heating in the ramp region does not manifest any anisotropy. It may be understood as there is intense wave activity in this region and the frequencies are in the ion acoustic range. The major effect of interaction of electrons with these low frequency waves consists in their angular scattering with a very weak energy change. This explains the absence of the anisotropy in this process. The temperatures (here the term "temperature" stands for conventional temperature determined as follows: $T = \frac{m}{3}\langle (v - \langle v \rangle)^2 \rangle >= \frac{m}{3n}\int f(v)(v - \langle v \rangle)^2 dv$, where $n$ is the number density of the corresponding population and $\langle ... \rangle$ means averaging over the distribution) of the core and reflected ion populations are presented in Figure 3d. The population separation shows that the temperature peak in the foot is associated only with the reflected ions and the peak values are lower than the estimation based on the full distribution gives. The peak temperatures in the foot are similar to the ion temperatures in the downstream (Figure 3d). Thus, the main heating of the core electron population happens on the shock ramp. Figure 3g shows the evolution of the ion anisotropy – the ion anisotropy, $A_i = T i_\perp / T i_\| - 1$, peaks at the ramp and can be unstable for generation of the ion cyclotron waves (the instability threshold is indicated with the red dashed curve). Intense electrostatic waves (with amplitudes up to 100 mV/m) are observed on the shock ramp (Figure 1g and Figure 3h). The waves' frequency range follows the structure of the background magnetic field (Figure 3i), and these waves' wave normal angles ($\theta$, the angle between the wave normal and the direction of the background magnetic field) are significantly oblique – up to 80 degrees (Figure 3j).

## The entropy dynamics during a bow shock crossing by Magnetospheric Multiscale (MMS) on November 2, 2017

According to e.g. Landau & Lifshitz (1977), for a state of a system close to a thermodynamic equilibrium, assumed to prevail at some distance upstream and downstream of the shock, the kinetic ensemble entropy $S = -k_B H$ per particle can be given by the expression (Parks et al. 2015):

$$H(\vec{r}) = -\frac{S(\vec{r})}{k_B} = -\int f(\vec{r}, \vec{v}) \cdot ln\big(f(\vec{r}, \vec{v})\big)d^3v,$$

where integral is over all velocity space, which can change but contains constant number of particles. We use here the entropy decomposition to the spatial and velocity entropy components established in (Liang et al. 2019) to proceed with the entropy density $H(\vec{r})$ in the spatial space. $f(\vec{r}, \vec{v})$ can be rewritten in terms of probability or normalized $f$ function, i.e. $f(\vec{r}, \vec{v}) = N\tilde{f}(\vec{r}, \vec{v})$ with $\int \tilde{f}(\vec{r}, \vec{v})d^3v = 1$ following (Parks et al. 2015), and will be used below for normalized solid angle distributions with the partial density distribution on energy (the binned energy channel for the spacecraft data, so that $N$ is the density of particles in the channel energy range, and $v^{*2}dv^*\int \tilde{f}(\vec{r}, \vec{v}^*) \sin\theta \, d\theta d\varphi = 1$, where $v^*$ is the velocity corresponding to the energy channel). The integral for the binned measurements constrained by the detector energy channels and then combined into the sky map (32x16 angles resolution in the MMS measurements) and the energy channels the integral becomes a sum decomposed to the energy channels ($i$ - index) and the solid angle bins ($i\theta, i\varphi$ indices):

$$H(\vec{r}) = -\sum_i \sum_{i_\theta, i_\varphi} n_i p_{i\theta i\varphi} \, ln \, n_i p_{i\theta i\varphi} v_i^2 \Delta v_i \, sin \, \theta_{i\theta} \, \Delta\theta \Delta\varphi =$$

$$= -\sum_i n_i \, ln \, n_i v_i^2 \Delta v_i - \sum_i n_i \sum_{i_\theta, i_\varphi} p_{i\theta i\varphi} \ln p_{i\theta i\varphi} \, sin \, \theta_{i\theta} \, \Delta\theta \Delta\varphi$$

$$H(\vec{r}) = -\sum_i n_i \, ln \, n_i v_i^2 \Delta v_i - \sum_i n_i h_i, \qquad (1)$$

where $p_{i\theta i\varphi}$ is the probability of the $\theta, \varphi$ state and $n_i$ is the partial density in the $i$-th energy channel (we mark the components as $A = \sum_i n_i \, ln \, n_i v_i^2 \Delta v_i$ and $B = \sum_i n_i h_i$ with corresponding indices $e$ and $p$ for electrons and protons respectively). We apply Eq. (1) to proceed the entropy changes across the bow shock bow shock crossing by Magnetospheric Multiscale (MMS) on November 2, 2017 discussed above. The entropy dynamics of the plasma volume going through the shock are shown in Figure 4 for electrons (blue) and ions (red). Figure 4a,c present the partial density values $n_i$ in the MMS energy channels ($N = \sum_i n_i$) of electrons and protons in the downstream and upstream of the shock (presented in Figures 1 and 3). The partial entropy values $h_i$ are shown in Figure 4 for electrons $h_{ei}$ (b) and protons $h_{pi}$ (d) estimated from the upstream and downstream distribution functions.

The increase of electron density across the shock is concentrated in the energy range from 10 to 1000 eV (Figure 4a). Electrons have almost uniform distribution at all energies, which is reflected in the high and almost equal values of $h_{ei}$ in all the energy channels (Figure 4b). This corresponds to much higher electron thermal speed values in comparison with the solar wind bulk velocity. Thus, $h_{ei}$ are close in the upstream and downstream regions, so, the changes in the distribution function do not lead to significant changes of the electron entropy values across the shock.

The proton population experiences the significant changes of the entropy across the shock. The lower values of the entropy corresponding to the solar wind flow of protons and helium in the upstream evolve to the smoothed distributions with almost 3 times higher values:

-in the upstream, the almost uniformly randomized fluxes in the wide range of energies below 100 eV and above 2 keV (the light red curve in Figure 4d) provide the highest values of the partial entropy values $h_{pi}$ of protons in the upstream. Conversely, the energy channels covering the directed solar wind flow (200 eV - 2 keV) have partial proton entropy values ~4-5 times lower, and these values determine the entropy density per particle in the upstream because of much higher statistical weights $n_{pi}$ (Fig4c, $A_p$ in Eq.(1));

-in the downstream, the less anisotropic proton flow distribution has higher partial entropy values $h_{pi}$ in all the energy channels (Fig4d) that together with spreading of the proton energy spectrum (the dark red curve in Fgi4c) is reflected in a 2-3 times increase of the partial entropy per particle.

The dynamics of the partial entropy $-H$ (Eq.(1)) is presented in Figure 4e for electrons and Figure 4f for protons. The contributions from the two components are shown with the dashed curve $A = \sum_i n_i \, ln \, n_i v_i^2 \Delta v_i$, and the dot curve ($B = \sum_i n_i h_i$). These components show the simultaneous increase of both the values $A_p$ nad $B_p$ for protons in the foot region (Figure 4f and Figure 4g, where the plasma density is shown in the background) but almost no changes for electrons (Figure 4e and 4g). In the foot region, the enhancement of entropy is provided by the reflected population. In the ramp, the partial entropy of the core population grows with the plasma temperature increase and the bulk flow velocity decrease. The electron entropy density is much higher in the upstream and has ~2-3% increase in the downstream (consistent with the results obtained for this shock crossing by Lindberg et al. (2022)), which is, however, about the estimation confidence interval, i.e. electron dynamics is almost adiabatic across this shock.

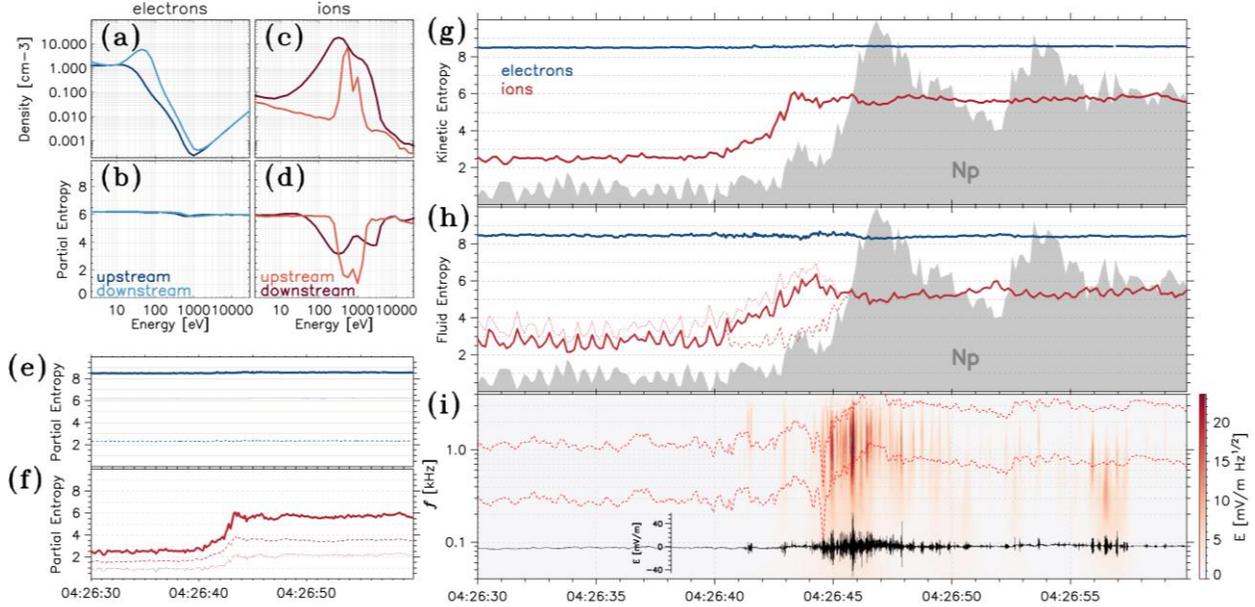

Figure 4. The partial (in the FPI energy channels) electrons density (a) and electrons partial entropy estimated by the Boltzmann kinetic approach in the upstream (light blue) and downstream (dark blue). The partial protons density (c) and protons partial entropy (d) in the upstream (light red) and downstream (dark red). The entropy dynamics for the proton (the red curve) and electron (the blue curve) populations estimated by the Boltzmann kinetic approach (g) and based on the thermodynamic potentials (h). The dot curve is based on the moments obtained from the full ion distribution function (spoiled by the reflected population) with a temperature enhancement in the foot region caused by the reflected population, i.e. one needs to process the core and reflected population moments separately. The dashed red curve represents the core population's entropy jump, which is coincident with the slowing down of the bulk velocity in the foot and the increase the ion temperature from the upstream to the downstream (Fig4c). The solid red curve represents the entropy dynamics based on the total pressure derived from partial pressures of the core and reflected populations: $p = (n_{cor}T_{cor} + n_{ref}T_{ref})$. The plasma density is shown with the grey background. The dynamics spectrum of wave electric field perturbations along the background magnetic field direction are shown in panel (k) with the waveform presented with the black curve (the waveform amplitude scale is in the right bottom).

The fluid (thermodynamic) description allows one to express the total entropy jump $\Delta S$ across the shock by assuming a smooth change from the upstream to the downstream quasi-equilibrium distribution function (see e.g. Serrin 1959; Landau & Lifshitz 1977). Then the jump in the entropy per unit mass at the shock, judged in the respective frames of the bulk plasma flow, is given by

$$\Delta S_{MHD} = ln\left[\frac{P_2}{P_1}\left(\frac{\rho_1}{\rho_2}\right)^{\gamma_a}\right] = k_B\,ln\frac{n_1}{n_2}\left(\frac{T_2}{T_1}\right)^{\frac{1}{\gamma_a-1}},$$

where $\gamma_a = 5/3$ is the adiabatic index suggesting that particles behave similarly to a monatomic ideal gas with three degrees of freedom. This value of $\gamma_a \cong 5/3$ was justified for the bow shock crossings by Lindburg et al. (2022). For this specific choice of $\gamma_a$, it follows that a gas, reacting strictly adiabatically at the shock compression, will not increase its thermal entropy. Figure 4h presents the thermodynamic approach to the entropy jump for ions and electrons confirming the results obtained from the kinetic approximation. A similar set of calculations using the measured electron distributions show that the electron entropy jump across the shock is much less: ~2-5%. This demonstrates that the main contribution to the entropy generation is provided by the core ions. Thus, tor this case, the downstream electron temperature $Te$ (where the electron anisotropy is almost zero, and marginal stability threshold for electron modes is not exceeded anywhere) matches the adiabatic expectation $Te/Te_0 = 2 - 2.1$ (here, $Te_0$ is the electron upstream temperature). The downstream proton temperature $T/Ti_0 \sim 6 - 6.5$ is much larger than the adiabatic expectation $T/Ti0 = 2$, so most of the entropy produced by the shock goes to the protons. In analogy to the so-called "magnetic pumping" mechanism (Lichko et al. 2017, Fowler et al. 2020), two basic ingredients are needed for plasma irreversible heating - the presence of a temperature anisotropy, induced by field amplification coupled to adiabatic invariance; and a mechanism to break the adiabatic invariance itself (the local generation of the ion cyclotron waves seen in Figure 4i). This confirms the numerical results reported by Gao et al. (2017, 2018), which highlight the anisotropy and wave generations to be the key factor for entropy growth across CS.

**Discussion**

The unprecedented high-time resolution plasma measurements aboard the four MMS spacecraft provide an opportunity for the detailed evaluation of the plasma dynamics across the front of the collisionless shock (CS) during multiple crossings of the terrestrial bow shock by MMS in 2015-2018. In the vicinity of the front one can distinguish several ion populations, well known from previous studies, namely: (1) the core ions; (2) the reflected ions, ions mirrored by the electrostatic potential and mirror force from the shock surface almost without energy change; (3) gyrating ions, the ions turning around the front several times and gaining energy due to gyrosurfing drift and acceleration; (4) the ions leaking from the downstream, these ions have a distribution similar to the partial downstream distribution (the part with the velocity component directed to the Sun), indicating that the source for this population is a leakage from the downstream.

Processing different populations separately allows one to evaluate the ion moments (bulk velocity, temperature, etc.) for each population. It shows that the change of the solar wind core population macroscopic parameters begins in the foot region but main change occurs in the ramp, where its bulk velocity slows down by the electrostatic potential and its temperature significantly increases. The foot region properties are determined by the counter-streaming plasma flows between upstream plasma and the reflected population. It is manifested by the formation of the temperature peak in the foot region (seen in Figure 1e) due to the relative motion of several instantaneously present ion species. The ion reflection at the front of a supercritical quasi-perpendicular shock is essentially non-specular as shown by (Gedalin 2016, Balikhin and Gedalin 2022). As the plasma beta is finite, the turning distances in the foot depend upon the phase of Larmor rotation during the interaction with the shock front and the velocity of a particular ion (Balikhin and Gedalin 2022). During the crossing of the foot region by the spacecraft the population of reflected ions varies as ions with shorter turning distances are continuously added to it. Results of ISEE mission have shown that the major ion thermalization occurs in the process of joint gyration of the directly transmitted and reflected ions downstream of the magnetic ramp (Sckopke et al. 1990). Overshoots and undershoots of the magnetic field downstream of the ramp are resulting from such a gyration due to the pressure balance. Ion heating takes place on a spatial scale that corresponds to the scale of the overshoots/undershoots trail and is significantly larger than the width of the magnetic ramp. The ion proxies based on the solar wind core population reproduce rather well the variations of the electrostatic potential (application of the total integral moments fails to reproduce the potential (Hanson et al. 2019)), so, the bulk velocity changes are well explained by the electrostatic potential drop on the ramp.

## Conclusions

The plasma entropy changes significantly from the upstream to the downstream, mostly due to the ions. The analysis of the entropy growth for different ion populations across the front of the stationary supercritical collisionless shock shows that the major growth of the total entropy is related to changes in the core ion population. It occurs in the foot region where the core and reflected ion populations co-exist and interact through ion cyclotron wave activity that results in modification of the core ion distribution. In addition, the entropy significantly increases in the ramp following the increase of the ion distribution anisotropy. The major dissipation mechanism consists in the heating of the core ion population. It operates in the foot and ramp regions, and the processes after the ramp play minor role in the evolution of the ion entropy. Little if any change of the electron entropy occurs, and if present, occurs in the vicinity of the ramp. Since current models of electron-ion anomalous resistivity would predict significant entropy increases in both the ions and electrons, these observations demonstrate that the role of the anomalous resistivity that was proposed in early works on collisionless shocks is negligible (Sagdeev 1966, 1960; Galeev 1976). In other words, "anomalous collisions" between electrons and ions caused by various plasma instabilities do not lead to the stochastisation of electron population and the irreversibility of the electron dynamics across the shock front. This is in full agreement with the current views that the electron thermalization at the shock front of a supercritical quasi-perpendicular shock is the result of the action of macroscopic magnetic and electric fields within the shock rather than anomalous collisions between electrons and ions due to various instabilities.

To summarize, the analysis of the entropy of ions and electrons populations across a stationary supercritical collisionless shocks shows that (1) the major growth of the total entropy is related to the core ion population thermalization; (2) ion thermalization starts in the foot (where the core and reflected ion populations co-exist and interact through wave activity that modifies ion core distribution) and finalizes in the ramp (where the core population entropy enhancement is maximal) with the increase of the ion distribution anisotropy; (3) the major dissipation mechanism consists of the anisotropic heating of the core ion population and its simultaneous scattering by the intense ion cyclotron waves; (4) the change of the electron entropy is significantly smaller if any; (5) this implies that the role of the anomalous resistivity that was proposed in early works on collisionless shocks is negligible.

## Acknowledgements

O.V.A and V.K. were supported by 80NSSC20K0697, and 80NSSC20K0697. O.V.A was supported by NSF grant number 1914670, NASA's Living with a Star (LWS) program (contract

80NSSC20K0218), and NASA grants contracts 80NNSC19K0848, 80NSSC22K0433, 80NSSC22K0522. MB acknowledges support by UK STFC grant ST/R000697/1